\documentclass[10pt,letterpaper]{article}
\usepackage[top=0.85in,left=2.75in,footskip=0.75in]{geometry}
 \usepackage{multirow}
\usepackage[pdftex]{graphicx}
\usepackage{subcaption}
\usepackage{enumitem}

\usepackage{amsmath,amssymb}

\usepackage{changepage}

\usepackage[utf8x]{inputenc}

\usepackage{textcomp,marvosym}

\usepackage{cite}

\usepackage{nameref,hyperref}

\usepackage{microtype}
\DisableLigatures[f]{encoding = *, family = * }

\usepackage[table]{xcolor}

\usepackage{array}
\usepackage{booktabs}

\newcolumntype{+}{!{\vrule width 2pt}}

\newlength\savedwidth

\raggedright
\usepackage[aboveskip=1pt,labelfont=bf,labelsep=period,justification=raggedright,singlelinecheck=off]{caption}

\makeatletter
\renewcommand{\@biblabel}[1]{\quad#1.}
\makeatother

\usepackage{lastpage,fancyhdr,graphicx}
\usepackage{epstopdf}
\fancyheadoffset[L]{2.25in}
\fancyfootoffset[L]{2.25in}
\newcounter{RZNumberOfComments}
\stepcounter{RZNumberOfComments}

\begin{document}
\vspace*{0.2in}

\begin{flushleft}
{\Large
\textbf\newline{Hate Speech Detection and Racial Bias Mitigation in Social Media based on BERT model}
}
\newline
\\
Marzieh Mozafari,
Reza Farahbakhsh,
No\"{e}l  Crespi
\\
\bigskip
\textbf CNRS UMR5157, T\'el\'ecom SudParis, Institut Polytechnique de Paris, \'Evry, France
\bigskip

%
%





* marzieh.mozafari@telecom-sudparis.eu

\end{flushleft}
\section*{Abstract}
THIS ARTICLE USES WORDS OR LANGUAGE THAT IS CONSIDERED PROFANE, VULGAR, OR OFFENSIVE BY SOME READERS.

Disparate biases associated with datasets and trained classifiers in hateful and abusive content identification tasks have raised many concerns recently. Although the problem of biased datasets on abusive language detection has been addressed more frequently, biases arising from trained classifiers have not yet been a matter of concern. In this paper, we first introduce a transfer learning approach for hate speech detection based on an existing pre-trained language model called BERT (Bidirectional Encoder Representations from Transformers) and evaluate the proposed model on two publicly available datasets that have been annotated for racism, sexism, hate or offensive content on Twitter. Next, we introduce a bias alleviation mechanism in hate speech detection task to mitigate the effect of bias in training set during the fine-tuning of our pre-trained BERT-based model. Toward that end, we use an existing regularization method to reweight input samples, thereby decreasing the effects of high correlated training set’ s $n$-grams with class labels, and then fine-tune our pre-trained BERT-based model with the new re-weighted samples. To evaluate our bias alleviation mechanism, we employed a cross-domain approach in which we use the trained classifiers on the aforementioned datasets to predict the labels of two new datasets from Twitter, AAE-aligned and White-aligned groups, which indicate tweets written in African-American English (AAE) and Standard American English (SAE), respectively. The results show the existence of systematic racial bias in trained classifiers, as they tend to assign tweets written in AAE from AAE-aligned group to negative classes such as racism, sexism, hate, and offensive more often than tweets written in SAE from White-aligned group. However, the racial bias in our classifiers reduces significantly after our bias alleviation mechanism is incorporated. This work could institute the first step towards debiasing hate speech and abusive language detection systems. 

\section*{Introduction}
\textbf{Disclaimer}: This article uses words or language that is considered profane, vulgar, or offensive by some readers. Owing to the topic studied in this article, quoting offensive language is academically justified but neither we nor PLOS in any way endorse the use of these words or the content of the quotes. Likewise, the quotes do not represent our opinions or the opinions of PLOS, and we condemn online harassment and offensive language.

Owning to the recent proliferation of user-generated textual contents in online social media, a wide variety of studies have been dedicated to investigating these contents in terms of hate or toxic speech, abusive or offensive languages, etc.,\cite{burnap2015, waseemhovy2016, Nobata2016, Davidson2017, snams2019, BadjatiyaG0V17}. With regard to the mobility and anonymous environment of online social media, suspect users, who generate abusive contents or organize the hate-based activities, exploit these online platforms to propagate hate and offensive contents towards other users and communities\cite{waseemhovy2016, Bertie2020}; where it leads to personal trauma, hate crime, cyber-bullying, and discrimination (mainly racial and sexual discriminations)\cite{park2018}. Therefore, online social media have been persuaded to define policies to remove such harmful content from their platforms since 2015\cite{Twitter, Facebook}.

The spread of hate speech and offensive language on online social media has received considerable attention from both academic and industrial environments to detect different types of hatred and toxicities (threats, obscenity, etc.). For example, different workshops and challenges such as the third Workshop on Abusive Language Online\cite{ALW3} and Kaggle’s Toxic Comment Classification Challenge\cite{kaggel} are conducted to address this issue by proposing different automated tools for identification of hate speech and abusive language on social media. 

Three main aspects of hate speech detection that rise to some challenges in this task are: 1) Definition of hate speech; 2) Designing and developing an automatic tool for identification of hate speech; 3) Tackling the problem of unintended data-driven and algorithm-driven biases in automatic hate speech detection tools; described as follows.

There is considerable disagreement about what exactly hate speech is\cite{plosone2019, Bertie2020}, and how different terms can be inferred as hatred or offensive in certain circumstances. For example, some terms such as ``n*gga'' and ``c*on'' were used to disparage African American communities, however, they were not known as offensive when used by peoples belonging to these communities\cite{sap2019}. In this study, we employ a commonly used definition of hate speech as any communication criticizing a person or a group based on some characteristics such as gender, sexual orientation, nationality, religion, race, etc., with or without using offensive or profane words.

To define automated methods with a promising performance for hate speech detection in social media, Natural Language Processing (NLP) has been used jointly with classic Machine Learning (ML)\cite{waseemhovy2016, Nobata2016, Davidson2017} and Deep Learning (DL) techniques\cite{mozafari2019, gamback2017, BadjatiyaG0V17}. The majority of contributions in classic supervised machine learning-based methods, for hate speech detection, rely on different text mining-based features or user-based and platform-based metadata\cite{Davidson2017, Waseem2018, Fortuna2018}, which require them to define an applicable feature extraction method and prevent them to generalize their approach to new datasets and platforms. However, recent advancements in deep neural networks and transfer learning approaches allow the research community to address these limitations. Although some deep neural network models such as Convolutional Neural Networks (CNNs)\cite{gamback2017}, Long Short-Term Memory Networks (LSTMs)\cite{BadjatiyaG0V17}, etc., have been employed to enhance the performance of hate speech detection tools, the requirement of a sufficient amount of labeled data and the inability of methods to be generalized have remained as open challenges. To address these limitations some transfer learning methods are proposed recently\cite{mozafari2019, Rizoiu2019}. However these methods enhanced the performance of hate speech detection models, they did not address the existing bias in data and algorithm.

From the bias's perspective, despite previous efforts into generating well-performed methods to detect hate speech and offensive language, the potential biases due to the collection and annotation process of data or training classifiers have raised a few concerns. Some studies ascertain the existence of bias regarding some identity terms (e.g., gay, bisexual, lesbian, Muslim, etc.) in the benchmark datasets and try to mitigate the bias using an unsupervised approach based on balancing the training set\cite{Dixon2018} or debiasing word embeddings and data augmentation\cite{park2018}. Moreover, some racial and dialectic bias exist in several widely used corpora annotated for hate speech and offensive language\cite{DavidsonBhattacharya2019, wiegand2019, sap2019}. Therefore, it is crucial to consider data-driven and algorithm-driven biases included in the hate speech detection system. Additionally, these kinds of race and gender discriminations caused by exciting biases in dataset or classifiers lead to unfairness against the same groups that the classifiers are trained to protect.

This study is an extended version of our previous work\cite{mozafari2019} at which we proposed a transfer learning approach for identification of hate speech in online social media by employing a combination of the unsupervised pre-trained model BERT\cite{bert2019} and new supervised fine-tuning strategies. Here, we investigate the effect of unintended bias in our pre-trained BERT-based model and use a generalization mechanism proposed by Schuster et.al \cite{Schuster2019}, for debiasing fact verification models, in training data by reweighting samples and then changing the fine-tuning strategies in terms of the loss function to mitigate the racial bias propagated through the model.

The main contributions of this work are as follows:
\begin{itemize}
\item Following our previous study\cite{mozafari2019}, we conduct a comprehensive experiment to inspect the impact of our transfer learning approach in a shortage of labeled data and in capturing syntactical and contextual information of all BERT transformers’ embeddings.
\item A regularization mechanism is used to mitigate data-driven and algorithm-driven bias by reweighting the training data and improving their generalization apart from their classes. We use two publicly available datasets for hate speech and offensive language detection. 
\item New fine-tuning strategy, in terms of the loss function, is employed to fine-tune the pre-trained BERT model by new re-weighted training data.
\item Finally, we perform a cross-domain validation approach to show the efficiency of the proposed bias mitigation mechanism.
\end{itemize}

\section*{Previous works}
In this section, we present in their respective subsections a comprehensive study of related works on hate speech detection, transfer learning, and data-driven and algorithm-driven bias analysis. Concerning these matters, we connect our work to the existing body of knowledge and convey our computational motivations.

\subsection*{Automatic hate speech detection} 
A majority of contributions have been provided towards the identification of hateful and abusive content in online social media\cite{Davidson2017, gamback2017, Olteanu2018, Ottoni2018, Mittos2019}. Applying a keyword-based approach is a fundamental method in hate speech detection task. Although using external sources such as the HateBase lexicon leads to a high-performing system in hate speech detection, maintaining and upgrading these resources are challenging\cite{plosone2019}. Furthermore, using specific hateful keywords in training data results in many false negatives related to the hateful samples, which are not containing those keywords\cite{Davidson2017, plosone2019}. Hence, we do not employ such external resources in this study. 

\noindent\textbf{Machine learning approach:}
To detect hateful and abusive contents, different machine learning approaches utilizing distinguishable feature engineering techniques have been employed in the literature\cite{Nobata2016, mehdad2016, waseemhovy2016}, and it is asserted that surface-level features such as a bag of words, word-level and character-level $n$-grams, etc., are the most predictive features in this task. Regarding classification perspective, different algorithms such as Na\"ive Bayes \cite{burnap2015}, Logistic Regression\cite{waseemhovy2016, Davidson2017}, Support Vector Machines\cite{Malmasi2018}, multi-view tacked Support Vector Machine (mSVM)\cite{plosone2019}, etc., have been used to train a classifier for predicting the hateful contents. 

As a baseline, Waseem et al.\cite{waseemhovy2016} addressed the problem of hate speech detection in Twitter by making a general definition of hateful content in social media based on guidelines inspired by Gender Studies and Critical Race Theory (CRT). Regarding that, they tried to annotate a corpus of 16,849 tweets as ``Racism'', ``Sexism'' and ``Neither'' by themselves, and the labels were inspected by “a 25-year-old woman studying gender studies and a non-activist feminist” for identifying potential sources of bias. To train their model, they used different sets of features such as word and character $n$-grams up to 4, gender, length, and location and investigated the impact of each feature on the classifier performance. Their results indicated that character $n$-grams are the most indicative features, and using location or length is detrimental. Furthermore, Davidson et al.\cite{Davidson2017} studied hateful and offensive contents in Twitter by sampling and annotating a $24K$ corpus of tweets as ``Hate'', ``Offensive'' and ``Neither''. They developed a variety of multi-class classifiers such as Logistic Regression, Na\"ive Bayes, Decision Trees, Random Forests, etc., on a set of features including Term Frequency–Inverse Document Frequency (TF-IDF) weighted $n$-grams, Part Of Speech (POS) tagging, sentiment scores, some tweet-level metadata such as the number of hashtags, mentions, retweets, URLs, etc. Although their results illustrated that Logistic Regression with L2 regularization performs the best in terms of accuracy, precision, and F1-scores, there are some social biases regarding anti-black racism and homophobia in their algorithm. Malmasi et al.\cite{Malmasi2018} proposed an ensemble-based system that used some linear SVM classifiers in parallel to distinguish hate speech from general profanity in social media. Recently, MacAvaney et al.\cite{plosone2019} discussed different aspects of an automatic hate speech system. They mainly addressed challenges pertaining to the definition of hate speech, dataset collecting and annotation process and its availability, and the characteristics of existing approaches. Furthermore, they proposed a multi-view tacked Support Vector Machine (mSVM) based approach that achieved near state-of-the-art performance; using word and character $n$-grams up to 5 as feature vectors. However, the issue of bias in data and trained models were not addressed there.

\noindent\textbf{Deep learning approach:}
Concerning the word representation as a dense vector pre-trained on a large amount of data, some basic deep learning approaches proposed to tackle the problem of hate speech\cite{gamback2017, Zhang2018}. The most frequently used word embeddings approaches are Word2Vec\cite{Mikolov2013}, Glove\cite{pennington2014} and FastText\cite{BojanowskiGJM16}.

As the first attempt, Djuric et al.\cite{Djuric2015} proposed a neural network-based model advantaging pagraph2vec embeddings to distinguish between hate speech and clean content. The proposed model incorporated two steps: in the first step, paragraph2vec embeddings were extracted from a continuous bag of words model, and in the second step, hateful and non-hateful contents were identified by applying a binary classifier counting on the extracted embeddings. Badjatiya et al.\cite{BadjatiyaG0V17}, who experimented on the dataset provided by Waseem and Hovy\cite{waseemhovy2016}, investigated three deep learning architectures: FastText, CNN, and LSTM. They used a combination of randomly initialized or GloVe-based embeddings with an LSTM neural network and a gradient boosting classifier. Their results outperformed the baseline from Waseem and Hovy\cite{waseemhovy2016}.

Different feature embeddings such as word embeddings and character $n$-grams were defined by Gambäck et al.\cite{gamback2017}, to solve the problem of identification of hate speech based on a CNN model. Afterward, a CNN+GRU (Gated Recurrent Unit network) neural network model was proposed by Zhang et al.\cite{Zhang2018} in which the model captured both word/character combinations (e. g., $n$-grams, phrases) and word/character dependencies (order information) with employing a pre-trained word2vec embeddings. Waseem et al.\cite{Waseem2018} brought a new insight to hate speech and abusive language detection tasks by proposing a multi-task learning framework to deal with datasets across different annotation schemes, labels, or geographic and cultural influences from data sampling. They proposed a transfer learning technique in which solving two hate speech detection tasks simultaneously and utilizing similarities between these two tasks leads to better generalization. Their experiments revealed that the multi-task learning framework produces better performance by switching between using a task as auxiliary and the other as primary. Using raw texts and domain-specific metadata from Twitter, Founta et al.\cite{Founta2019} proposed a unified classification model at which different types of abusive language such as cyberbullying, hate, sarcasm, etc., were efficiently performed.

\subsection*{Transfer learning}
In the machine learning domain, transfer learning is a concept in which prior knowledge gained from one domain and task will be applied to solve another problem from a different domain and task but related one somehow. In NLP tasks, the word embeddings models that encode and represent an entity such as word, sentence, document, etc., to a fixed-length vector, were the first attempts toward applying the transfer learning approach to adjust to the best performance. Using pre-trained word embeddings such as Word2Vec\cite{Mikolov2013}, Glove\cite{pennington2014}, and FastText\cite{BojanowskiGJM16} exploited from a large text corpus such as Wikipedia, news articles, etc., result in great advances in different NLP tasks especially for problems at which there may not be enough training data. However, these pre-trained models suffer for their disability to better disambiguate between the correct sense of a given word regarding different contexts in which it appears. To address this issue, different contextual-based pre-trained models such as Universal Language Model Fine-Tuning (ULMFiT)\cite{Ruder2018}, Embedding from Language Models (ELMO)\cite{Matthew_2018}, OpenAI’ s Generative Pre-trained Transformer (GPT)\cite{Radford2018}, and Google’s BERT model\cite{bert2019} emerged. In these models, a universal language model is pre-trained on a general-domain corpus by applying different techniques such as bi-directional LSTM\cite{Matthew_2018}, unidirectional transformer\cite{Radford2018}, and bidirectional transformer\cite{bert2019} and then a downstream task will be fine-tuned using discriminative methods.

For the first time, Waseem et al.\cite{Waseem2018} applied a multi-task learning strategy as a transfer learning model to transfer knowledge between two different hateful and offensive datasets. Their results indicated the ability of multi-task learning to generalize to new datasets and distributions in hate speech detection tasks. Afterward, using a combination of GloVe and pre-trained ELMO words embeddings, Rizoiu et al.\cite{Rizoiu2019} proposed a transfer learning approach for hate speech and abusive language detection (two datasets provided by\cite{waseemhovy2016,  Davidson2017}). To adjust the ELMo representation to the hate speech detection domain, they applied a bi-LSTM layer independently trained left-to-right and right-to-left on both tasks simultaneously and then extracted sentence embedding using a max-pooling approach. At the end, a specific classifier was trained for each task. Due to the jointly solving both tasks, the insights learned from one task can be transferred to the other task. Comparing the results from these two transfer learning-based studies indicates that the approach of Waseem et al.\cite{Waseem2018} outperforms Rizoiu et al.\cite{Rizoiu2019}, therefore, we consider the approach of Waseem et al.\cite{Waseem2018} as our baseline here and compare our proposed method with that.

Due to the lack of undoubted labeled data and the inability of surface features to capture the subtle semantics in text, identification of hateful and offensive content is an intricate task\cite{Malmasi2018}. To address this issue, we use the pre-trained language model BERT for hate speech classification and try to fine-tune a specific task by leveraging information from different transformer encoders. 

\subsection*{Bias detection and mitigation in hate speech systems}
Recently the great efforts have taken to examine the issue of data bias in hate speech and offensive language detection tasks. Dixon et al.\cite{Dixon2018} confirmed the existence of unintended bias between texts containing general identity terms (e.g. lesbian, gay, Islam, feminist, etc.) and a specific toxicity category; attributed to the disproportionate representation of texts containing certain identity terms through different categories in training data from Wikipedia Talk pages dataset. Therefore, they tried to quantify and mitigate this form of unintended bias by expanding training and test datasets under some generalization strategies for identity terms. Following some debiasing methods (Debiased Word Embeddings, Gender Swap, and Bias fine-tuning), Park et al.\cite{park2018} tried to measure and debias gender bias in abusive language detection system. Afterward, Wiegand et al.\cite{wiegand2019} conveyed that unintended biases in datasets are not just restricted to the identity terms and gender, and they are by cause of focused data sampling approaches. Consequently, the high classification scores on these datasets, mainly containing implicit abuse, are due to the modeling of the bias in those datasets. Datasets containing biased words resulted from biased sampling procedure cause a huge amount of false positives when testing on other datasets. They showed that some query words used for sampling data from Tweeter that are not correlated with abusive tweets but are included in tweets with sexist or racist remarks are biased as well. For example, query words such as commentator, sport, and gamergate used by Waseem et al.\cite{waseemhovy2016} to sample data from Twitter, are not correlated with Sexism class but are one of the most frequent words in this category. Furthermore, Badjatiya et al.\cite{Badjatiya2019www} proposed a two-step bias detection and mitigation approach. At first, various heuristics were described to quantify the bias and a set of words in which the classifier stereotypes were identified. Then, they tried to mitigate the bias by leveraging knowledge-based generalization strategies in training data. The results show that their approach can alleviate the bias without reducing the model performance significantly.

Recently, Davidson et al.\cite{DavidsonBhattacharya2019} and Sap et al.\cite{sap2019} investigated the racial bias against African American English (AAE) dialects versus Standard American English (SAE) in the benchmark datasets with toxic content, especially from the Twitter platform. They declared that the classifiers trained on these datasets tend to predict contents written in AAE as abusive with strong probability. Furthermore, Sap et al.\cite{sap2019} introduced a way of mitigating annotator bias through dialect, but they did not mitigate the bias of the trained model. 

We propose a pre-trained BERT-based model to address the problem of hate speech detection and the data-driven and algorithm-driven biases, which extends the prior literature in two significant ways. First, it outperforms previous methods in terms of F1-measure by applying different fine-tuning strategies and employing different syntactic and semantic information embedded in different layers of BERT. Second, it addresses unintended bias in data or trained models and tries to mitigate the racial bias in our pre-trained BERT-based classifiers. Our bias mitigation approach is close to what Davidson et al.\cite{DavidsonBhattacharya2019} did at which they just addressed the racial bias in the benchmark hate speech datasets. However, in this study, we use a bias mitigation mechanism to alleviate racial bias included in datasets and trained classifiers by leveraging a regularization mechanism in training set proposed by Schuster et. al.\cite{Schuster2019} for alleviating the bias in fact verification tasks.
 
\section*{Materials and methods}
\begin{figure}[h]
\centering
\begin{adjustwidth}{-1.5in}{0in}
\includegraphics[width=1.0\textwidth]{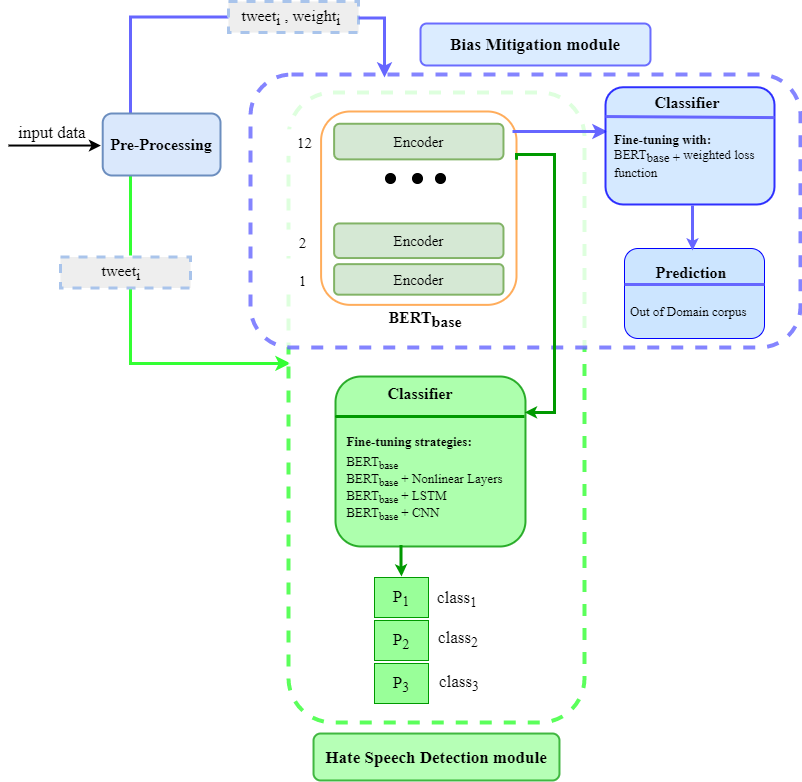}
\caption{\textbf{The proposed framework for hate speech detection and bias mitigation tasks}. It consists of two different modules: Hate Speech Detection and Bias Mitigation with different inputs as a result of different pre-processing approaches. The pre-trained BERT\textsubscript{base} is a common component between two modules that is fine-tuned differently in respect of each module’s goal.}
\label{fig:framework}
\end{adjustwidth}
\end{figure}

In this section, we introduce our proposed framework for hate speech detection and unintentional bias analysis and mitigation. As shown in Fig~\ref{fig:framework}, our approach contains two main modules: \textbf{(1) Hate Speech Detection module} and \textbf{(2) Bias Mitigation module}; where the pre-trained BERT\textsubscript{BASE} component is shared between two modules. Here, we describe and analyze more deeply the hate speech detection module, proposed in our previous study\cite{mozafari2019}, and then the details related to the proposed bias mitigation mechanism will be provided in Section Bias mitigation module.

\subsection*{BERT-based hate speech detection module}
According to Fig~\ref{fig:framework}, given tweets in the training set as input data, the pure texts of them are extracted from the pre-processing component regarding a set of specific rules, described in the related subsection. Then, the processed tweets are fed into the pre-trained BERT model to be fine-tuned according to different strategies with task-specific modifications. At the end, using the trained classifiers we predict the labels of the test set and evaluate the results.

To analyze the ability of the BERT transformer model on the identification of hate speech, we describe the mechanism used in the pre-trained BERT model at first. BERT is a multi-layer bidirectional transformer encoder trained on the English Wikipedia and the Book Corpus containing 2,500M and 800M tokens, respectively, and it has two models named BERT\textsubscript{BASE} and BERT\textsubscript{LARGE} detailed as follows:

\begin{itemize}
  \item  \textbf{BERT\textsubscript{BASE}}: contains 12 layers (transformer blocks), 12 self-attention heads, and 110 million parameters.
  \item  \textbf{BERT\textsubscript{LARGE}}: contains 24 layers, 16 attention heads, and 340 million parameters.
\end{itemize}

Each of BERT\textsubscript{BASE} and BERT\textsubscript{LARGE} has two versions: uncased and cased. The uncased version has only lowercase letters. In this study, we use the uncased version of the pre-trained BERT\textsubscript{BASE} model. A sequence of tokens, as a pre-processed sentence, in maximum length 512 is fed to the BERT model as input. Then two segments are added to each sequence as [CLS] and [SEP] by BERT tokenizer. [CLS] embedding, which is the first token of the input sequence, is used as a classification token since it contains specific classification information in each layer. The [SEP] token, an artifact of two-sentence tasks, separates segments and we will not use it in our classification because we have only single-sentence inputs. As the output, BERT produces a 768-dimensional vector to represent each input sequence.

\subsubsection*{Fine-tuning strategies}
As we are dealing with textual content from social media in our task and the BERT model is pre-trained on general corpora, it is crucial to analyze the contextual information extracted from pre-trained BERT's transformer layers. Different levels of syntactic and semantic information are encoded in different layers of the BERT model, and according to\cite{bert2019} the lower layers of the BERT model may contain information that is more general whereas the higher layers contain task-specific information. Hence, we need to fine-tune it on our hate speech detection task with annotated datasets. Here, fine-tuning means to train and update the entire pre-trained BERT model along with the additional untrained classifier layers of 768 dimensions (considering different fine-tuning strategies) on top of the pre-trained BERT\textsubscript{BASE} transformer (more information about these transformer encoders' architectures are presented in\cite{bert2019}). In the following, a brief description of different fine-tuning strategies, explained in detail in our previous study\cite{mozafari2019}, are included.

\textbf{BERT based fine-tuning:}
To fine-tune BERT with this strategy, we use the output of the [CLS] token, a vector of length 768, from 12th transformer encoder and feed it as input to a fully connected neural network without hidden layer. To classify each input sample a softmax activation function is employed to the hidden layer.

\textbf{Insert nonlinear layers:}
Similar to the previous strategy, the output of the [CLS] token, a vector of length 768, from the latest transformer encoder is used as an input to a fully connected neural network with two hidden layers in size 768. Leaky Relu activation function with negative slope = 0.01 is applied on two hidden layers and, at the end, a softmax activation function for the final layer is used.

\textbf{Insert Bi-LSTM layer:}
Contrary to the previous strategies, all outputs of the latest transformer encoder are fed to a bidirectional recurrent neural network (bi-LSTM) on the top of the BERT model. The final hidden state is directed to a fully connected neural network with a softmax activation function to do the classification operation.

\textbf{Insert CNN layer:}
Rather than using the output of the latest transformer encoder, here we use the outputs of all transformer encoders in the BERT model as an input to a convolutional neural network with a window size: (3 and hidden size of BERT which is 768 in BERT\textsubscript{BASE} model). Then, by applying a MaxPooling method on the convolution's outputs, the maximum values of each transformer encoder are extracted, and a vector is generated to be fed as input to a fully connected neural network. In the end, the classification function is performed by applying a softmax activation function.

\section*{Experiment setup}
This section presents details about the datasets and the pre-processing step used for the identification of hate speech. Furthermore, we provide some technical details related to the implementation part at the end of the section.
\subsection*{Dataset description}
In this study, we experiment with three publicly available datasets widely studied on Twitter provided by Waseem and Hovy\cite{waseemhovy2016}, Waseem\cite{waseem2016} and Davidson et al.\cite{Davidson2017}, which are detailed in the following:
 
\textbf{Waseem and Hovy\cite{waseemhovy2016}/Waseem\cite{waseem2016}}: Within two months period, Waseem and Hovy\cite{waseemhovy2016} collected 136,052 tweets from Twitter and, after some filtering, annotated a corpus containing 16,914 tweets as ``Racism", ``Sexism" and ``Neither".
First using an initial ad-hoc approach, they tried to search common slurs and terms related to religious, sexual, gender, and ethnic minorities. Secondly, from the first results, they identified the most frequent terms in tweets containing hate speech. For example, hashtag ``\#MKR” which was related to a public Australian TV show, My Kitchen Rules, and caused many sexist tweets directed at the female participants. At the end to make their sampling process more general, they crawled more tweets containing clearly abusive words and potentially abusive words but they are not abusive in context, as negative sampling. The final collected corpus ($16K$) was annotated by experts and ascertained by a 25 years old woman studying gender studies and non-activist feminist to reduce annotator bias. Waseem\cite{waseem2016} also provided another dataset to investigate the impact of expert and amateur annotators on the performance of classifiers trained for hate speech detection. Therefore, they collected 6,909 tweets for hate speech and annotated them as ``Racism", ``Sexism", ``Neither" and ``Both" by amateurs from CrowdFlower crowdsourcing platform and experts having a theoretical and applied knowledge of the abusive language and hate speech. Their efforts result in a set of 4,033 tweets where there was an overlap of 2,876 tweets between their new dataset and the one provided by Waseem and Hovy\cite{waseemhovy2016}. Since both datasets are overlapped partially and they used the same strategy in definition of hateful content, we merged these two datasets following Waseem et al.\cite{Waseem2018} to make our imbalance data a bit larger (we followed all the rules provided in Section 3.2 of Waseem et al.\cite{Waseem2018} paper to merge two datasets. For more details, please refer to that paper). In the rest of the paper, we refer to this aggregated dataset as \textbf{Weseem-dataset}.

\textbf{Davidson et al.\cite{Davidson2017}}: Employing a set of particular terms from a pre-defined lexicon of hate speech words and phrases, called HateBase\cite{hatebase}, Davidson et al.\cite{Davidson2017} crawled 84.4 million tweets from 33,458 twitter users. To annotate collected tweets as ``Hate", ``Offensive" or ``Neither", they randomly sampled $25k$ tweets and asked users of CrowdFlower crowdsourcing platform to label them. After labeling each tweet by annotators, if their agreement was low, the tweet was eliminated from the sampled data. In the rest of the paper, we refer to this dataset as \textbf{Davidson-dataset}.

Table \ref{tab:data_description} shows a brief description of class distribution in both datasets.

\begin{table}[h]
\begin{adjustwidth}{-01.25in}{0in}
\caption{\textbf{Datasets description}. The columns show the total number of tweets, the different categories and the percentage of tweets belong to each one in the datasets, respectively.}
\label{tab:data_description}
\begin{tabular}{@{}llc@{}}
\hline
\textbf{Dataset} &\textbf{ \#Tweets} &\multicolumn{1}{l} {\textbf{Classes and  percentage of membership}} \\ \hline
\multirow{3}{*}{Waseem-dataset \cite{waseemhovy2016}\cite{waseem2016}} & \multirow{3}{*}{19697} & Racism (10.73\%) \\ \cmidrule(l){3-3} 
 &  & Sexism (21.15\%) \\ \cmidrule(l){3-3} 
 &  & Neither (68.12\%) \\ \hline
\multirow{3}{*}{Davidson-dataset \cite{Davidson2017}} & \multirow{3}{*}{24783} & Hate (5.77\%) \\ \cmidrule(l){3-3} 
 &  & Offensive (77.43\%) \\ \cmidrule(l){3-3} 
 &  & Neither (16.80\%) \\ \hline
\end{tabular}
\end{adjustwidth}
\end{table}

\subsection*{Pre-processing}
For simplicity and generality, we consider the following criteria in order to filter the raw dataset and make it clean as the input of our model:

\begin{itemize}[noitemsep]
  \item Converting all tweets to lower case.
  \item Removing mentions of users, for the sake of protecting the user’s identities.
  \item Removing embedded URLs in tweets’ content.
  \item Removing common emoticons, because in this study we do not consider emotions in our analysis.
  \item Identifying elongated words and converting them into short and standard format; for example, converting ``yeeeessss” to ``yes”.
  \item Removing hashtag signs (\#) and replacing the hashtag texts by their textual counterparts, where there is not any space between them; for example, we convert hashtag ``\#notsexist" to ``not sexist".
  \item Removing all punctuation marks, unknown uni-codes and extra delimiting characters
  \item Keeping all stop words, because our model trains the sequence of words in a text directly.
  \item Eliminating tweets with a length of less than 2 after applying all aforementioned pre-processing steps.
\end{itemize}

\subsection*{Implementation}
\label{implementation}
Our hate speech detection and bias mitigation modules are implemented with publicly available pytorch-pretrained-bert library\cite{pytorch}. We utilize the pre-trained BERT model, text tokenizer, and pre-trained WordPiece provided in the library to prepare the input sequences and train the model. Using BERT tokenizer, we tokenize each tweet (as input sentence) in such a way that invalid characters are removed and all the words are lowercased. Following the original BERT\cite{bert2019}, words are split to subword by employing WordPiece tokenization. Due to the shortness of input sentences' length, the maximum sequence length is set to 64 and in any case of shorter or longer length, it will be padded with zero values or truncated to the maximum length, respectively. We train our classifiers with different fine-tuning strategies with a batch size of 32 for 3 epochs on Google Colaboratory tool\cite{googlecolab} with an NVIDIA Tesla K80 GPU and 12G RAM; as the implementation environment. During training, we use an Adam optimizer with a learning rate of 2e-5 to minimize the Cross-Entropy loss function; furthermore, the dropout probability is set to 0.1 for all layers.

\subsection*{Evaluation metrics}
In general, classifiers with higher precision and recall scores are preferred in classification tasks. However, due to the imbalanced classes in the hate speech detection datasets, we tend to make a trade-off between these two measures. Therefor, we summarize models’ performance into macro averaged F1-measure, which is the geometric mean of precision and recall and gives more insights into the performance characteristics of each classification model.

\section*{Experiment results}
Here, we investigate the impact of using a pre-trained BERT-based model with different fine-tuning strategies on the hate speech detection task. Additionally, we show different aspects of our transfer learning-based approach by analyzing the proposed model deeply. 

To train the model, we need to split Waseem-dataset and Davidson-dataset into training, validation and test sets. Considering the disparate distribution of tweets in different classes described in Table \ref{tab:data_description}, it is justifiable that we are dealing with imbalanced datasets (to adjust the classes’ distribution of the datasets, we do not oversample or undersample the datasets because hate speech and offensive languages are real phenomena and we want to provide the datasets to the classifiers as realistic as possible). Using a stratified sampling technique 0.8, 0.1 and 0.1 portions of tweets in each class: Racism, Sexism, and Neither or Hate, Offensive, and Neither are selected for training, validation, and test sets in each dataset, respectively.

We consider models proposed by Davidson et al.\cite{Davidson2017} and Waseem et al.\cite{Waseem2018} as our baselines in which a classic method and a deep neural network model are created respectively. To do so, following the original work\cite{Davidson2017}, we create an SVM classification method proposed by the authors and we train a machine learning model using a multi-task learning framework proposed by Waseem et al.\cite{Waseem2018}. In addition to these two baselines, we compare our results with the methods proposed in\cite{waseemhovy2016, Zhang2018, Park2017, plosone2019} on the corresponding datasets. Using two hate speech datasets, we examine the performance of our model, with different fine-tuning strategies, in contrast to the baselines and state-of-the-art approaches. The evaluation results on the test sets are reported in terms of macro averaged F1-measure in Table \ref{bert-fine-tune}. The differences between the results provided in Table \ref{bert-fine-tune} and what were reported in the original works are due to we implemented some models and report macro averaged F1-measures.

\begin{table}[h]
\begin{adjustwidth}{-01.25in}{0in}
\caption{\textbf{Performance evaluation.} Performance of different trained classifiers on Waseem-dataset and Davidson-dataset in terms of F1-measure are reported in \subref{perf_waseem} and \subref{perf_davidson}, respectively.}
\label{bert-fine-tune}
\begin{subtable}{0.6\textwidth}
\begin{tabular}{@{}lc@{}}
\toprule
\textbf{Model} & \multicolumn{1}{l}{\textbf{F1-Measure}} \\ \midrule
Waseem and Hovy\cite{waseemhovy2016} & 75 \\
Waseem et al.\cite{Waseem2018} & 80 \\
Zhang et al.\cite{Zhang2018} & 82 \\
Park et al.\cite{Park2017} & 83 \\ \midrule
BERT\textsubscript{BASE} & 81 \\
BERT\textsubscript{BASE} + Nonlinear Layers & 76 \\
BERT\textsubscript{BASE} + bi-LSTM & 86 \\
BERT\textsubscript{BASE} + CNN & 88 \\ \bottomrule
\end{tabular}
\caption{Performance evaluation on Waseem-dataset.}
\label{perf_waseem}
\end{subtable}
\begin{subtable}{0.55\textwidth}
\flushright
\begin{tabular}{@{}lc@{}}
\toprule
\textbf{Model} & \multicolumn{1}{l}{\textbf{F1-Measure}} \\ \midrule

Davidson et al.\cite{Davidson2017} & 84 \\
Zhang et al.\cite{Zhang2018} & 94 \\
Waseem et al.\cite{Waseem2018} & 89 \\ 
MacAvaney et al.\cite{plosone2019} & 77 \\ \midrule
BERT\textsubscript{BASE} & 91 \\
BERT\textsubscript{BASE} + Nonlinear Layers & 87 \\
BERT\textsubscript{BASE} + bi-LSTM & 92 \\
BERT\textsubscript{BASE} + CNN & 92 \\ \bottomrule

\end{tabular}
\caption{Performance evaluation on Davidson-dataset.}
\label{perf_davidson}
\end{subtable}
\end{adjustwidth}
\end{table}

Table \ref{bert-fine-tune} shows that, in both datasets, all the BERT-based fine-tuning strategies except BERT + nonlinear classifier on top of it outperform the existing approaches or they achieve competitive results. According to Table \ref{perf_waseem}, on Waseem-dataset, the highest F1-measure value is achieved by BERT\textsubscript{BASE} + CNN which is 88\% and there is a 5\% improvement from the best performance achieved by Park et al.\cite{Park2017} method. In addition, applying different models on Davison-dataset, reported in Table \ref{perf_davidson}, also confirms the previous observation and shows that using the pre-trained BERT model as initial embeddings and fine-tuning the model with a CNN yields the best performance in terms of F1-measure; where it is 92\%. On Davidson-dataset, comparing the best F1-measure value achieved by BERT\textsubscript{BASE} + CNN model with the best-performed model proposed by Zhang et al.\cite{Zhang2018} indicates that our model achieved a 2\% decrease in performance than\cite{Zhang2018}; where the F1-measure is 94\%. We posit that this is due to the fact that Zhang et al.\cite{Zhang2018} have merged the Hate and Offensive classes of Davidson-dataset together and solved the problem of hate speech detection as a binary classification which it made the task more simplified counter to our specific multi-class classification approach.

From deep learning neural network perspective, according to the literature\cite{kim2014}, CNN works well with data that have a spatial relationship. In hate speech classification tasks, there is an order relationship between words in a document and CNN learns to recognize patterns across space. In the combination of BERT + CNN, although convolutions and pooling operations lose information about the local order of words, it has already captured by BERT encoders and its position embeddings in different layers. On the other hand, from the language modeling perspective, BERT + CNN uses all the information included in different layers of pre-trained BERT during the fine-tuning phase. This information contains both syntactical and contextual features coming from lower layers to higher layers of BERT. Therefore, this model works the best of all models tested.

\subsubsection*{Performance evaluation with a limited amount of training data}
\begin{figure}[h]
			\centering
			\begin{adjustwidth}{-1.75in}{0in}
				\begin{subfigure}[b]{0.6\textwidth}
				      \includegraphics[width=1.2\textwidth]{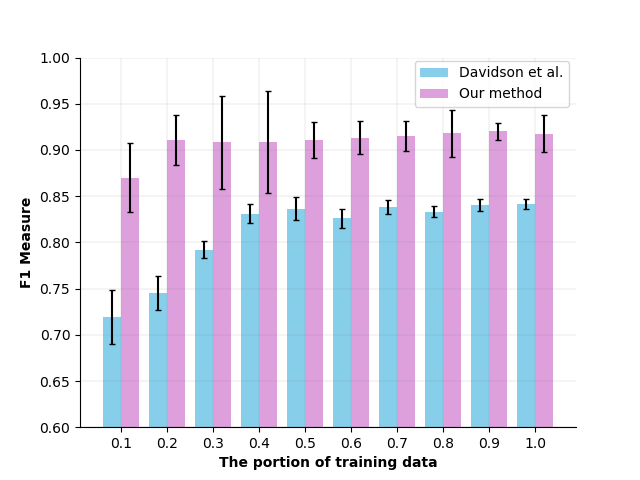}
                                                \caption{\centering  Davidson-dataset}
                                                  \label{fig:d_prop}
                                           \end{subfigure}
  				 \hfill
				\hspace*{\fill}
				 \begin{subfigure}[b]{0.6\textwidth}
 				         \includegraphics[width=1.2\textwidth]{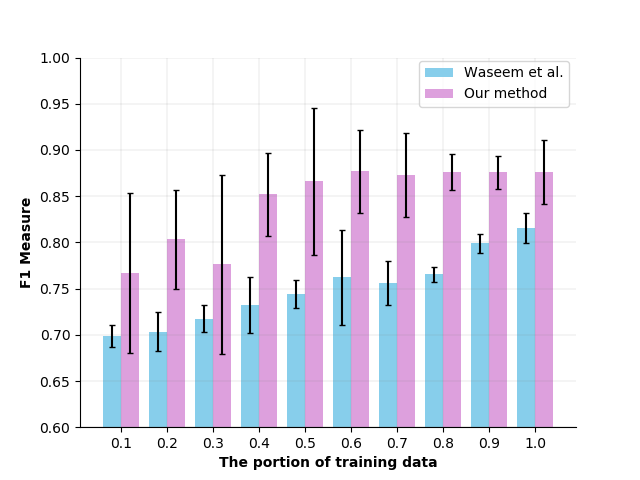}
                                                   \caption{\centering Waseem-dataset}
                                                   \label{fig:w_prop}
                                            \end{subfigure}
			           \caption{\textbf{The performances of hate speech detection models trained with a variation of training sets on Davidson and Waseem datasets}. The x-axis is the portion of the training and validation sets used for training our BERT-based model and the baselines, the y-axis shows the F1-measure.}
				\label{fig:prop_analysis}
		             \end{adjustwidth}
		\end{figure}

In common practice the more the fraction of training set is, the higher the performance of algorithms will be. One advantage of leveraging the pre-trained model is to be able to train a model for downstream tasks within a small training set. Due to the lack of a sufficient amount of labeled data in some classification tasks, mainly hate speech detection here, using the pre-trained BERT model can be effective. We inquire into the performance of hate speech detection models in terms of F1-measure when the amount of labeled data is restricted. Fig~\ref{fig:prop_analysis} shows the evaluation results of the baselines and our pre-trained BERT-based model on different portions of training examples, over a certain concentration range $[0.1 - 1.0]$. We train and test each model 10 times and report the results in terms of their mean and standard deviation. For each dataset, we select training and test sets according to the description included in Section Experiment results. We do not use the validation set (10\% of the dataset) for Davidson et al.\cite{Davidson2017} baseline model but it is used in Waseem et al.\cite{Waseem2018} baseline. In Waseem et al.\cite{Waseem2018} baseline model we are dealing with a multi-task learning approach, therefore in each iteration, the training and validation sets of a specific task which is going to be trained are selected. For our proposed method, we report the performance of the pre-trained BERT model fine-tuned with inserting a CNN layer on top of it; the best performing fine-tuning strategy. To see how the models perform on different portions of training and validation sets, we restrain the amount of training and validation sets in such a way that only a specific portion of them are available for the models during the training.

The experiment results demonstrate that our pre-trained BERT-based model brings a significant improvement to small size data and has comparable performance on different portions of training data in comparison to the baseline models. According to Fig~\ref{fig:d_prop}, the smallest portion of training data, which is 0.1, used in the training phase of our model is able to yield the F1-measure of almost 87\% where it is 72\% for Davidson baseline. By increasing the portion of training data, the performance of the Davidson baseline gradually increases up to 83\% (where the portion of the training set is 0.5) and then remains considerably stable, whereas the performance of our model does not significantly improve. This finding supports the theory that using a pre-trained BERT-based model causes a decrease in the size of the required training data to achieve a specific performance. From Fig~\ref{fig:w_prop},
we can observe that the performance of the multi-task learning approach proposed by Waseem et al.\cite{Waseem2018} gradually increases and it depends on the portion of training data. However, the performance of our model is mostly stable during the growth of training data, especially by including more than 0.3 of training data.

\subsubsection*{BERT embeddings analysis}		
To see how informative different 12 layers of transformer encoders of the BERT model are, we extract embeddings for each sentence in our datasets, from pre-trained BERT model before and after fine-tuning. Here, we use the uncased BERT\textsubscript{BASE} model with 12 transformer blocks, 12 attention heads, and a hidden layer size of 768. For this purpose, we use an online service called bert-as-service\cite{bertasservice} to map a variable-length sentence into a fixed-length vector representation and extract sentence embeddings from different layers of the BERT model.

\begin{figure}
\centering
\begin{adjustwidth}{-1.25in}{0in}
\begin{subfigure}[b]{1\textwidth}
   \includegraphics[width=1.1\linewidth]{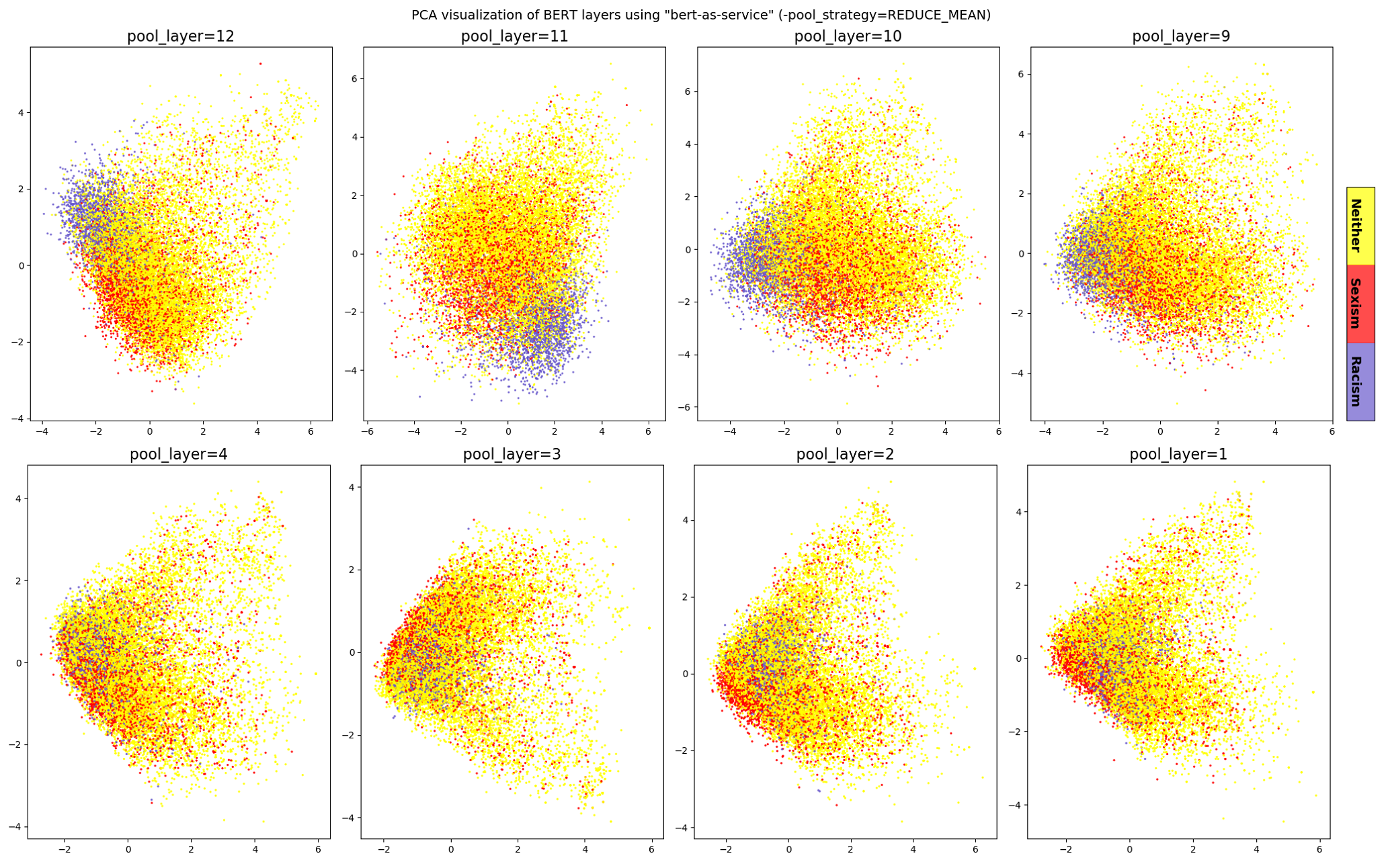}
  \caption{\centering Before fine-tuning}
   \label{fig:z_initial} 
\end{subfigure}

\begin{subfigure}[b]{1\textwidth}
  \includegraphics[width=1.1\linewidth]{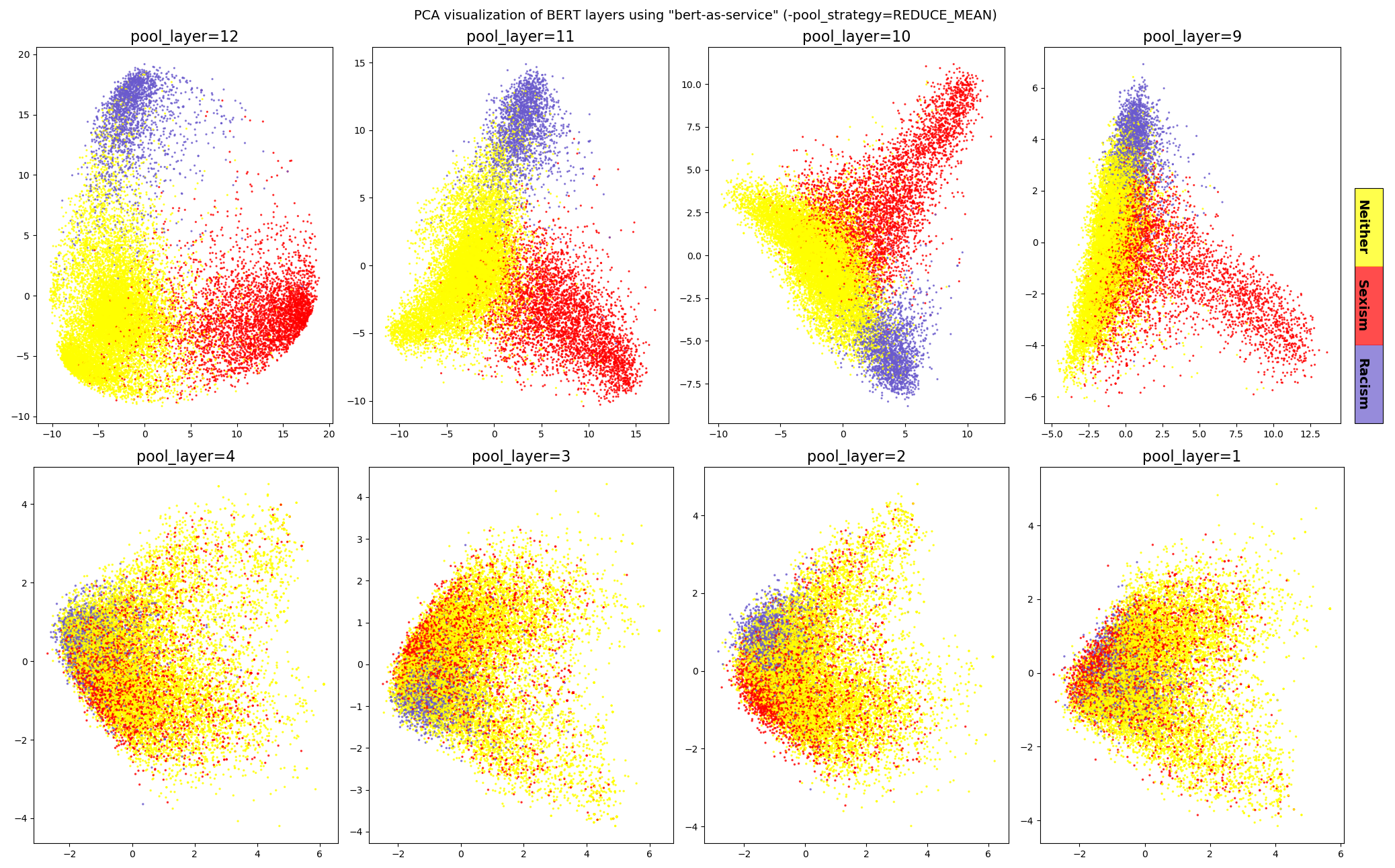}
  \caption{\centering After fine-tuning}
   \label{fig:z_enrich}
\end{subfigure}
\caption[Two numerical solutions]{\textbf{Waseem-samples' embeddings analysis before and after fine-tuning.} To investigate the impact of information included in different layers of BERT, sentence embeddings are extracted from all the layers of the pre-trained BERT model  fine-tuning, using the bert-as-service tool. Embedding vectors of size 768 are visualized to a two-dimensional visualization of the space of all Waseem-dataset samples using PCA method. For sake of clarity, we just include visualization of the first 4 layers (1-4), which are close to the training output, and the last 4 layers (9-12), which are close to the word embedding, of the pre-trained BERT model before and after fine-tuning.}
\label{fig:z_embedding_analysis}
\end{adjustwidth}
\end{figure}

We extract the vector representation of all samples in Davidson and Waseem datasets separately from the original pre-trained BERT model and the one we fine-tuned on our downstream tasks. Each sample is translated into a 768-dimensional vector. As [CLS] special token appeared at the start of each sentence does not have richly contextual information before fine-tuning the model on a specific classification task, we take all the tokens’ embeddings in a sentence and apply a REDUCE-MEAN pooling strategy to get a fixed representation of a sentence. Given the sentence representations from the pre-trained BERT model before and after fine-tuning, Principal Component Analysis (PCA) builds a mapping of 768-dimensional vector’s representation to a 2D space shown in Fig~\ref{fig:z_embedding_analysis} for Waseem-dataset. There are three classes of the data, illustrated in purple, red, and yellow corresponding to Racism, Sexism, and Neither classes, respectively. 

Sentence Embeddings from the first 4 layers (1-4) and the last 4 layers (9-12) of pre-trained BERT model before fine-tuning on Waseem-dataset are represented in Fig~\ref{fig:z_initial}.
Regarding the fact that different pre-trained BERT layers capture different information, we can see that sentences’ representation from each class in the first 4 layers is highly sparse which means the Euclidean pairwise distance between sentences in each class is large in the high dimensional space. However, the sentence embeddings in the last 4 layers are a bit more clustered in comparison to the first 4 layers according to the class which they belong to; Especially for racism samples. This observation is on the grounds that, pre-trained BERT model is trained on Wikipedia and Book Corpus data and encodes enough prior knowledge of the general and formal language into the model. However, this knowledge is not specific to a particular domain; here hate speech contents form social media with informal language. Therefore, before fine-tuning the model on our task different layers of BERT cannot capture the contextual and semantic information of samples in each class and cannot congregate similar sentences in a specific class.

After fine-tuning our model, on Waseem-dataset, with BERT\textsubscript{BASE} + CNN strategy, which performs as the best fine-tuning strategy on both datasets, we can observe in Fig~\ref{fig:z_enrich}
that the model captures contextual information in which racism, sexism, and neither content exist and clusters samples strongly tight in the last 4 layers. It causes the high-performance evaluation result using this fine-tuning strategy in our previous study\cite{mozafari2019}. The same result is yielded by Davidson-dataset's embeddings visualization included in \nameref{S1_Fig}.

\newpage
\subsubsection*{Error analysis}
As we observed in Experiment result section, although we have very interesting results in terms of F1-measure, it is needed to examine how the model predicts false positives and false negatives. To understand better this phenomenon, in this section we perform an analysis of the error of the model. We investigate the test datasets and their confusion matrices resulted from the BERT\textsubscript{BASE} + CNN model as the best fine-tuning approach; depicted in Tabel \ref{tab:table2}. According to Tabel \ref{conf_waseem} for Waseem-dataset, it is obvious that the model can separate sexism from racism content properly. Only two samples belonging to racism class are misclassified as sexism and none of the sexism samples are misclassified as racism. A large majority of the errors come from misclassifying hateful categories (racism and sexism) as hatless (neither) and vice versa. 0.9\% and 18.5\% of all racism samples are misclassified as sexism and neither respectively whereas it is 0\% and 12.7\% for sexism samples. Almost 12\% of neither samples are misclassified as racism or sexism. As Tabel \ref{conf_davidson} makes clear for Davidson-dataset, the majority of errors are related to hate class where the model misclassified hate content as offensive in 63\% of the cases. However, 2.6\% and 7.9\% of offensive and neither samples are misclassified respectively.

\begin{table}[h]
\begin{adjustwidth}{-01.25in}{0in}
\caption{Confusion matrix of the both Waseem-dataset (\subref{conf_waseem}) and Davidson-dataset (\subref{conf_davidson}).}
\label{tab:table2}
\begin{subtable}{0.6\textwidth}
\begin{tabular}{@{}l|ccc@{}}
\toprule
\multirow{2}{*}{\textbf{Label}} & \multicolumn{3}{c|}{\textbf{Predicted}}                                               \\
                                & \multicolumn{1}{l}{Racism} & \multicolumn{1}{l}{Sexism} & \multicolumn{1}{l}{Neither} \\ \midrule
Racism                          & 169                        & 2                          & 39                          \\
Sexism                          & 0                          & 362                        & 53                          \\
Neither                         & 133                        & 22                         & 1160                        \\ \bottomrule
\end{tabular}
\caption{Waseem-dataset's confusion matrix}
\label{conf_waseem}
\end{subtable}
\begin{subtable}{0.55\textwidth}
\flushright
\begin{tabular}{@{}l|ccc@{}}
\toprule
\multirow{2}{*}{\textbf{Label}} & \multicolumn{3}{c|}{\textbf{Predicted}}                                                \\
                                & \multicolumn{1}{l}{Hate} & \multicolumn{1}{l}{Offensive} & \multicolumn{1}{l}{Neither} \\ \midrule
Hate                            & 42                       & 90                            & 10                          \\
Offensive                       & 29                       & 1867                          & 25                          \\
Neither                         & 4                        & 29                            & 382                         \\ \bottomrule
\end{tabular}
\caption{\centering {Davidson-dataset's confusion matrix}}
\label{conf_davidson}
\end{subtable}
\end{adjustwidth}
\end{table}

Our manual inspection on a subset of data showed that, in Davidson-dataset, the model has more tendency to base predictions on certain words such as ``n*gga'', ``b*tch'', etc., due to the imbalance dataset (Hate:5.77\% and Offensive:77.43\%). Furthermore, in some cases containing implicit abuse (like subtle insults) such as: 

\textit{Tweet: @user: Some black guy at my school asked if there were colored printers in the library. "It’s 2014 man you can use any printer you want I said.}

our model cannot capture the hateful content and therefore misclassifies. It should be noticed that even for a human it is difficult to discriminate against this kind of implicit abuses.

According to the strategy used in collecting data in Davidson-dataset, some tweets with specific language (written within the African American Vernacular English) and geographic restriction (United States of America) are oversampled and result in high rates of misclassification \cite{DavidsonBhattacharya2019, Waseem2018}. However, these misclassifications do not confirm the low performance of our classifier because annotators tended to annotate many samples containing disrespectful words as hate or offensive without any presumption about the social context of tweeters such as the speakers’ identity and dialect or surrounding context of the tweet; whereas they were just offensive or even neither tweets such as:

\textit{Tweet: @user: If you claim Macklemore is your favorite rapper I’m also assuming you watch the WNBA on your free time fagg*t.}

\textit{Tweet: @user: @user typical c*on activity.}

These kinds of tweets are some samples containing offensive words and slurs that are not hateful or offensive in all cases, and writers of them used this type of language in their daily communications, but they were labeled as hate by annotators without considering the context.

\section*{Bias mitigation module}
As depicted in Fig~\ref{fig:framework}, our proposed framework consists of two main modules. This section concentrates on the bias mitigation module at which we address the problem of data-driven and algorithm-driven biases in hate speech detection. We explore existence bias in the datasets and then try to mitigate the bias in the proposed pre-trained BERT-based model by applying a generalization mechanism.

\label{bias_analysis_module}
\subsection*{Towards unbiased training data}
Although a lot of effort has been done in proposing and developing a real-world abusive language and hate speech detection systems, their potential biases due to the collecting and annotating process of data or training classifiers on them have raised a few concerns. Recently, some studies tried to address this issue. As demonstrated in \cite{DavidsonBhattacharya2019, wiegand2019, sap2019} there is some racial and dialectic bias in several widely used corpora annotated for toxic language (e.g., hate speech, abusive speech, or other offensive speech). 

To the best of our knowledge, it is the first time that we are addressing bias mitigation through trained classifier rather than data sampling and annotation process. Here, we try to improve the generalization in the existence of the racial and dialect bias by using a generalization mechanism in the training data. To mitigate the bias propagated through the models on which the benchmark datasets are trained, we leverage a re-weighting mechanism, by inspiring from the recent work of Schuster et al.\cite{Schuster2019}. First, we assess the explicit bias in the datasets and investigate phrases in training set causing it. Then, we reweight the samples in training and validation sets to make smooth the correlation between the phrases in training samples and the classes to which they belong. After optimizing the bias in the training set, we acquire re-weighted scores for each sample and feed our pre-trained BERT-based model with new training and validation sets (as depicted in Fig~\ref{fig:framework}, where tweets and corresponding weights are as an input of the Bias Mitigation module). During the fine-tuning, the loss function of the classifier will be updated with re-weighted scores to alleviate the existing bias in training samples. 

The high classification scores in hate speech detection and offensive language systems are likely due to modeling the bias from training datasets. Therefore, we assess the explicit bias in Davidson and Waseem datasets and investigate phrases in training sets causing it. To do so, the $n$-gram distribution in training and test sets is inspected and the high frequently $n$-grams, that are extremely correlated with a particular class, are extracted. We use the Local Mutual Information (LMI)\cite{evert2005} to extract high frequently $n$-grams in each class. For any given $n$-gram $w$ and class $c$, LMI between $w$ and $c$ is defined as follows:

\begin{equation}
 \label{eq:lmi}
LMI(w,c)=p( w, c).log(\frac{p( c| w)}{p ( c )} )
\end{equation}

where $p(c|w)$ and $p(c)$ are calculated by $\frac{count(w,c)}{count(w)}$ and $\frac{count(c)}{|D|}$, respectively. Furthermore, $p(c)$ and  $p(w|c)$ are calculated by $\frac{count(c)}{|D|}$ and $\frac{count(w,c)}{|D|}$, respectively. $|D|=$ is the number of occurrences of all $n$-grams in the training set.

\begin{figure}[h]
\centering
\begin{adjustwidth}{-2.25in}{0in}
\begin{subfigure}[b]{0.6\textwidth}
\includegraphics[width=1.28\textwidth]{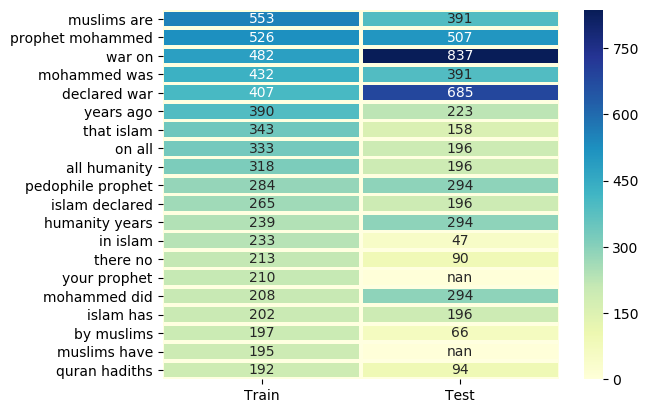}
\caption{\centering  Racism class}
\label{fig:racism}
\end{subfigure}
\hfill
\hspace*{\fill}
 \hspace{19mm}
\begin{subfigure}[b]{0.6\textwidth}
\includegraphics[width=1.28\textwidth]{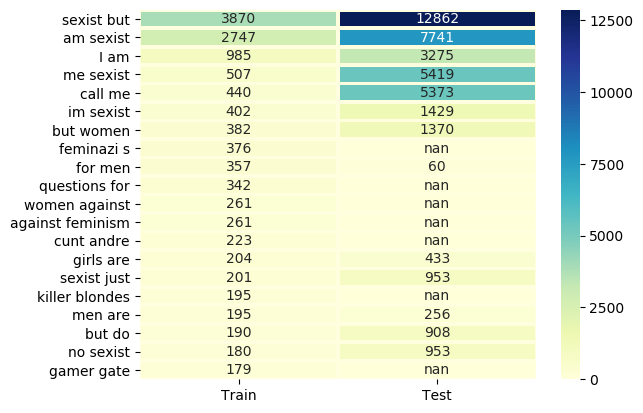}
\caption{\centering Sexism class}
\label{fig:sexism}
\end{subfigure}
\caption{\textbf{The top 20 LMI-ranked $n$-grams ($n=2$) that are highly correlated with the negative classes of Waseem-dataset (Racism and Sexism) in the training and test sets.} nan value denotes computationally infeasible, as the occurrence is zero in the test set.}
\label{fig:bigram_waseem}
\end{adjustwidth}
\end{figure}

\begin{figure}[h]
\centering
\begin{adjustwidth}{-2.25in}{0in}
\begin{subfigure}[b]{0.6\textwidth}
\includegraphics[width=1.28\textwidth]{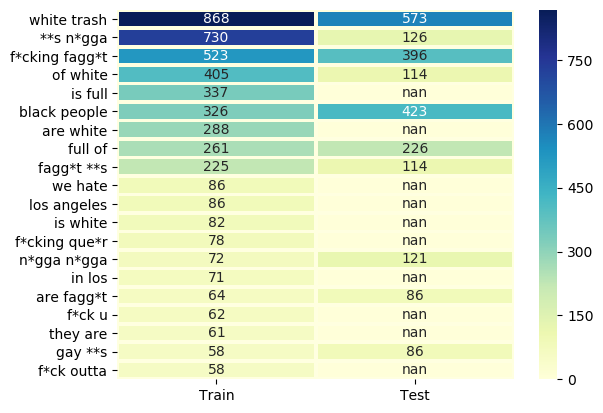}
\caption{\centering Hate class}
\label{fig:hate}
\end{subfigure}
\hfill
\hspace*{\fill}
 \hspace{19mm}
\begin{subfigure}[b]{0.6\textwidth}
\includegraphics[width=1.28\textwidth]{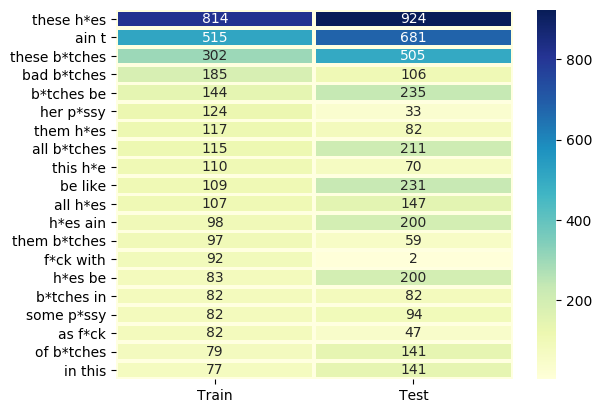}
\caption{\centering Offensive class}
\label{fig:offensive}
\end{subfigure}
\caption{\textbf{The top 20 LMI-ranked $n$-grams ($n=2$) that are highly correlated with the negative classes of Davidson dataset (Hate and Offensive) in the training and test sets.} nan value denotes computationally infeasible, as the occurrence is zero in the test set.}
\label{fig:bigram_davidson}
\end{adjustwidth}
\end{figure}

Figs~\ref{fig:bigram_waseem} and ~\ref{fig:bigram_davidson} exhibit the 20 top LMI-ranked $n$-grams ($n=2$) that are highly correlated with the Racism and Sexism classes of Waseem-dataset and Hate and Offensive classes of Davidson-dataset in the training and test sets, respectively. Using training and test data, a heat map with legend color bar, column and row side annotations is generated in Figs~\ref{fig:racism} and ~\ref{fig:sexism} for Racism and Sexism and Figs~\ref{fig:hate} and ~\ref{fig:offensive} for Hate and Offensive classes. The legend color bar indicates the correlation between LMI values and colors, and the colors are balanced to ensure the light yellow color represents zero value. LMI values indicate with $LMI.10^{-6}$. Illustrating the most frequently 2-grams in Racism class in Fig~\ref{fig:racism} shows that tweets in this class are containing some domain-specific expressions such as ``islam'' and ``muslims'' at which they are likely to be associated with Racism class (as hateful class). On the other hand, in Fig~\ref{fig:sexism} some general keywords such as ``women'', ``feminism'', and ``sexist'' are highly associated with Sexism class. These kinds of correlations are true for both training and test sets’ samples except some phrases in which there is not any occurrence in the test set and is indicated as nan value. Therefore, it is perceived that there are some idiosyncrasies in the dataset construction for each class and they are described as stereotype bias in the rest of the paper.

The same stereotype bias exists in Hate and Offensive classes of Davidson-dataset (Fig~\ref{fig:bigram_davidson}) where samples containing specific terms such as ``n*gga'', ``fagg*t'', ``que*r'', etc., are highly correlated with Hate class. On the other hand, the samples containing terms such as ``h*es'' and ``b*tch'' are associated with Offensive class. This kind of stereotype bias can be transferred to the classifier during the training process and creates a tendency for predicting new samples containing this stereotype as a negative class.

\subsection*{Re-weighting mechanism} 
This section presents the mechanism to alleviate the bias in our hate speech detection model. We describe how samples belonging to each class are assigned a positive weight according to their correlation with the different classes. After that, samples with new weights are fed to our pre-trained BERT-based model. To mitigate the bias initiated by $n$-grams high correlated to each class in our proposed model, we use an algorithm introduced by Schuster et al.\cite{Schuster2019}, for debiasing a fact verification model, to reweight the samples. We believe that it is the first attempt to reduce the systematic bias existing in hate speech datasets with such kind of re-weighting mechanism.

Bias made by high frequently 2-grams per class in training and validation sets can be constrained by defining a positive weight $\alpha^{i}$  for each sample  $x ^{i}$, tweet in training and validation sets, in such a way that the importance of tweets with different labels containing these phrases are increased. Considering each sample as $x ^{i}$, its label as $y ^{i}$ and each 2-gram in training set as $w_{j}$, we define a bias toward each class $c$ using Eq~\ref{eq:re-weights} \cite{Schuster2019}. 

\begin{equation}
 \label{eq:re-weights}
{b_{j}}^{c}=\frac{\sum_{i=1}^{n}I_{[w_{j}^{(i)}]}(1+\alpha ^{(i)})I_{[y^{(i)}=c]}}{\sum_{i=1}^{n}I_{[w_{j}^{(i)}]}(1+\alpha ^{(i)})}
\end{equation}

Where $I_{[w_{j}^{(i)}]}$ and $I_{[y^{(i)}=c]}$ are the indicators for $w_{j}$ to be in tweet $x^{i}$ and lable $y^{i}$ to be in class $c$.

To find balancing weights $\alpha$ that result in the minimum bias  \cite{Schuster2019}, we have to solve an optimization problem as follows:

\begin{equation}
 \label{eq:alpha}
min(\sum_{j=1}^{|V|}max_{c}(b_{j}^{c})+\lambda ||\overrightarrow{\alpha}||_{2})
\end{equation}

It should be noted that we acquire $\alpha$ values in the pre-processing step and before feeding training and validation sets to our BERT-based model. To integrate the weights associated with each sample into our model, the loss function of our pre-trained BERT-based classification model has to be changed. In our previous study\cite{mozafari2019} we used Cross-entropy loss function\cite{bishop2006} as a loss function when optimizing our classification model on top of the pre-trained BERT model. However, in this study, we change the loss function in such a way that it includes weights as well.

Let $y = y_{1} , . . . , y_{n}$ be a vector representing the distribution over the classes $1, . . . , n$, and let $\widehat{y} = \widehat{y_{1}}, . . . , \widehat{y_{n}}$ be the classifier output .The categorical cross entropy loss measures the dissimilarity between the true label distribution $y$ and the predicted label distribution $\widehat{y}$, and is defined as cross entropy as follows:
\begin{equation}
\label{eq:original_loss_function}
\text{Loss}_{\mathit{cross-entropy}} (\widehat{y},  y) = -\sum_{i=1}^{n} y_{i} \log ( \widehat{y})
\end{equation}

While for the re-weighted approach, the training objective is reweighted from the Eq~\ref{eq:original_loss_function} to:

\begin{equation}
 \label{eq:weighted_loss_function}
\text{Weighted-Loss}_{\mathit{cross-entropy}} (\widehat{y},  y) = -\sum_{i=1}^{n}  (1+\alpha ^{(i)}) y_{i} \log ( \widehat{y})
\end{equation}

\subsection*{Scrutinizing bias mitigation mechanism}
To further analyze the impact of the regularization mechanism through training and validation sets and reweighting the samples for bias mitigation, we investigate how the models trained on samples with and without weights predict on new datasets (cross-domain data). We use a dataset collected from twitter by Blodgett et al.\cite{blodget2016} including a demographically associated dialectal language named African American English (AAE), known as Black English, which is a dialect of American English spoken by millions of black people across the United States. They exploited a set of geo-located tweets by leveraging a distantly supervised mapping between authors and the demographics of the place in which they live. They filtered out 16 billion collected tweets in such a way that tweets geo-located with coordinates that matched a U.S. Census blockgroup remained; which contains 59.2 million publicly available tweets. Consequently, four different demographic categories of non-Hispanic whites, non-Hispanic blacks, Hispanics, and Asians are created using the information about population ethnicity and race from the U.S. Census. They proposed a probabilistic mixed-membership language model to learn demographically aligned language models for each of the four demographic categories utilizing words associated with particular demographics. At the end, they calculated a posterior proportion of language from each category in each tweet. Following Davidson et al.\cite{DavidsonBhattacharya2019} recent work, to analysis racial bias propagated with the pre-trained BERT-based model with and without the re-weighting mechanism, we define two categories of tweets as follows:

\textbf{AAE-aligned}: filtering the tweets with the average posterior proportion greater than 0.80 for the non-Hispanic black category and less than 0.10 for Hispanic + Asian together to address the African Americal English language (AAE).

\textbf{White-aligned}: filtering the tweets with the average posterior proportion greater than 0.80 for the non-Hispanic white category and less than 0.10 for Hispanic + Asian together to address the Standard American English (SAE).

After filtering out the tweets not satisfying the above conditions, we result in a set of 14.5m and 1.1m tweets written in non-Hispanic white (White-aligned) and non-Hispanic black (AAE-aligned) languages, respectively. These two new categories show the racial alignment of the language that their authors used. In the following, we explain how we use these datasets to evaluate our pre-trained BERT-based classifier with and without re-weighting mechanism to alleviate racial bias.

\textit{\textbf{Research Question}}: Our research question here is that, whether or not our BERT-based classifiers trained on Waseem and Davidson datasets with and without the re-weighting mechanism, have any preference in assigning tweets from AAE-aligned and White-aligned categories to a negative class (Racism, Sexism, Hate or Offensive). If it is yes, how our proposed bias alleviation mechanism reduces this tendency.

Considering each tweet $t$ in AAE-aligned dataset as $t_{\textit{black}}$ and in White-aligned dataset as $t_{\textit{white}}$, we define two hypotheses $H1$ and $H2$ for each class $c_{i}$ where $c_{i}=1$ denotes membership of $t$ in class $i$ and $c_{i}=0$ in the opposite. Therefore, $H1$ is equivalent to $P(c_{i}=1|black)=p(c_{i}=1|white)$ in which the probability of $t$ to be a member of a negative class $i$ is independent of the racial group at which it belongs to. $H2$ is equivalent to $P(c_{i}=1|black)>p(c_{i}=1|white)$ or $P(c_{i}=1|black)<p(c_{i}=1|white)$ in which the probability of $t$ to be a member of a negative class $i$ is dependent on the racial group at which it belongs to.

To assess our hypotheses, we conduct an experiment in which we sample 10000 tweets from each AAE-aligned and White-aligned groups and feed them as a test set to our pre-trained BERT-based classifiers trained on Davidson and Waseem datasets, separately, with and without the re-weighting mechanism to predict the membership probability of each tweet in each class. For each classifier, trained on Waseem and Davidson datasets, we create a vector containing the membership probability $p_{i}$ of each class $i$ in size of the number of samples in each group (10000). Indeed, we obtain one vector per each class $i$ for tweets in two AAE-aligned and White-aligned groups and calculate the portion of tweets assigned to each class $i$ for each group as follows:

$\widehat{p_{i_{black}}}=\frac{1}{n}\sum_{j=1}^{n}p_{ij}$ where $j$ denotes the samples from AAE-aligned and $\widehat{p_{i_{white}}}=\frac{1}{n}\sum_{j=1}^{n}p_{ij}$ where $j$ denotes the samples from White-aligned and $n=10000$. To examine the racial bias tendency of each classifier on each class $i$, we also calculate $\frac{\widehat{p_{i_{black}}}}{\widehat{p_{i_{white}}}}$ as an indicator. If this portion is greater than 1 then it indicates that our classifier has a higher propensity to assign AAE-aligned tweets to a specific class $i$ rather than White-aligned tweets.

To see how significant the differences between $\widehat{p_{i_{black}}}$ and $\widehat{p_{i_{white}}}$ are, we apply an independent samples t-test between two groups which results in $t$ and $p$ values, where $t$ indicates the difference between two groups and the difference within the groups and $p$ indicates the probability that the results from the tweets samples occurred by chance. A low value of $p$ shows that our membership probabilities assigned with the classifiers did not occur by chance (Here, the $p$ values for all the classes are less than $0.001$ which indicated as *** in Table \ref{bias_analysis_1}). 

All the results are shown in Table \ref{bias_analysis_1}, where we computed the aforementioned statistics with and without including the bias alleviation mechanism in our pre-trained BERT-based models trained on different datasets. Statistics signed with $*$ indicate the values after debiasing the training sets. For fine-tuning the pre-trained BERT model, we have tried all fine-tuning strategies, but report the results from the best performing strategy in bias mitigation task which is BERT\textsubscript{BASE} fine-tuning strategy. The first row shows the performance of classifier trained on Waseem dataset on two-race groups before and after reweighting. The second row indicates the same results for Davidson dataset. In all cases, the tweets belonging to AAE-aligned group are more frequently predicted as a member of negative classes than White-aligned which indicates existing of systematic bias in two datasets.
\begin{table}[h]
\begin{adjustwidth}{-2.25in}{0in} 
\centering
\caption{\textbf{Racial bias analysis before and after reweighting the training data.} To quantify the impact of the re-weighting mechanism in alleviating the racial bias propagated through trained classifiers, we examine our BERT-based classifiers trained on Waseem and Davidson datasets with and without re-weighting mechanism on AAE-aligned and SAE-aligned samples.}
\label{bias_analysis_1}
\begin{tabular}{@{}lllllll|lllll@{}}
\cmidrule(l){3-12}
                                &                              & \multicolumn{5}{c|}{\textbf{Before reweighting}}                        & \multicolumn{5}{c}{\textbf{After reweighting}}   \\ \midrule
\textbf{Dataset}                         & \multicolumn{1}{l|}{\textbf{Class}}     & $\widehat{p_{i_{black}}}$ & $\widehat{p_{i_{white}}}$ & \textit{t} & \textit{p}   & $\frac{\widehat{p_{i_{black}}}}{\widehat{p_{i_{white}}}}$  & $\widehat{p_{i_{black}}}^{*}$ & $\widehat{p_{i_{black}}}^{*}$ & $t^{*}$      & $p^{*}$   & $\frac{\widehat{p_{i_{black}}}}{\widehat{p_{i_{white}}}}^{*}$ \\ \midrule
\multirow{2}{*}{Waseem-dataset} & \multicolumn{1}{l|}{Racism}    & 0.049            & 0.005            & 10.450     & *** & 10.593 & 0.028   & 0.007   & 6.852   & *** & 3.726 \\
                                & \multicolumn{1}{l|}{Sexism}    & 0.162            & 0.055            & 31.715     & *** & 2.923  & 0.235   & 0.092   & 15.949  & *** & 2.561 \\ \midrule
\multirow{2}{*}{Davidson-dataset}       & \multicolumn{1}{l|}{Hate}      & 0.058            & 0.026            & 84.986     & *** & 2.230  & 0.043   & 0.031   & 1.815   & *** & 1.384 \\
                                & \multicolumn{1}{l|}{Offensive} & 0.360            & 0.143            & 17.913     & *** & 2.515  & 0.193   & 0.106   & 120.607 & *** & 1.823 \\ \bottomrule
\end{tabular}\\
We just consider negative classes and ``Neither'' class in both datasets is excluded.
\end{adjustwidth}
\end{table}

Surprisingly, there is a significant difference across AAE-aligned and White-aligned groups in Racism class’ s estimated rates. Our classifier on Waseem-dataset classifies tweets in AAE-aligned group as Racism 10.5 times more probably than White-aligned without reweighting, which indicates potential bias carried with our trained model and not dataset itself. However, after applying bias alleviation mechanism by reweighting the samples and decreasing the correlation between high frequently 2-grams and each negative class, we can observe that our model decreases $\frac{\widehat{p_{i_{black}}}}{\widehat{p_{i_{white}}}}^{*}$ by 6.8 times for Racism class. This kind of racial bias reduction is true for Sexism class as well.

For Davidson-dataset, we observe that tweets in AAE-aligned are classified as Hate and Offensive more frequently than White-aligned. The classifier trained on Davidson-dataset before applying the re-weighting mechanism gives Hate label to AAE-aligned tweets with 5.8\% and to White-aligned tweets with 2.6\%, as opposed to 4.3\% and 3.1\% in re-weighted classifier. Consequently, $\frac{\widehat{p_{i_{black}}}}{\widehat{p_{i_{white}}}}^{*}$ gets down by 0.85 times in comparison with $\frac{\widehat{p_{i_{black}}}}{\widehat{p_{i_{white}}}}$ in Hate class. For Offensive class, the bias mitigation rate is 0.70 where the probability of assigning AAE-aligned samples to Offensive class reduces from 36\% to 19\%. Comparing results for Hate and Offensive classes shows that the classifiers trained on Davidson-dataset classify AAE-aligned tweets more frequently as Offensive rather than Hate; which is the result of the unbalanced dataset we used to train the classifiers.

From Table \ref{bias_analysis_1} it is inferred that substantial racial bias perseveres even after using our bias alleviation mechanism, however, it is generally reduced for cases in which classifiers are trained with re-weighted samples. It means that still, our re-weighted classifiers favor assigning tweets from AAE-aligned more probably to negative classes rather than White-aligned after bias mitigation. Given our cross-domain approach for evaluating the bias mitigation mechanism, we hypothesize that differences between Davidson and Waseem datasets’ keywords and language and AAE-aligned and White-aligned languages, which are not included in our bias mitigation mechanism, lead classifiers to classify tweets written by African-Americans (AAE-aligned group) as negative classes excessively.

We investigate the performance of the pre-trained BERT-based model (with BERT\textsubscript{BASE} strategy for fine-tuning) after applying the proposed re-weighting mechanism on the in-domain dataset as well; where test data come from Waseem-dataset and Davidson-dataset. Performance evaluation of the classifier before and after reweighting is showed in Table \ref{bias_analysis_2} in terms of macro precision, recall, and F1-measure. 

\begin{table}[h]
\caption{\textbf{Performance evaluation after applying the re-weighting mechanism.} To quantify the impact of the re-weighting mechanism in the performance of our pre-trained BERT-based model (with BERT\textsubscript{BASE} strategy for fine-tuning), we examine the classifier trained on Waseem and Davidson datasets with and without re-weighting mechanism on the training set  in terms of macro precision, recall, and F1-measure.}

\label{bias_analysis_2}
\begin{tabular}{@{}l|ccc|ccc@{}}
\toprule
\textbf{}        & \multicolumn{3}{c|}{\textbf{Before reweighting}} & \multicolumn{3}{c|}{\textbf{After reweighting}} \\ \midrule
\textbf{Dataset} & Precision       & Recall       & F1-measure       & Precision       & Recall       & F1-measure      \\ \midrule
Wassem-dataset   & 81              & 81           & 81               & 76               & 79            & 78               \\ \midrule
Davidson-dataset & 91              & 90           & 91               & 85               & 88            & 86               \\ \bottomrule
\end{tabular}
\end{table}

According to Table \ref{bias_analysis_2}, reweighting the training data has a negative effect on the performance of our classifier in detecting Racism, Sexism, Hate, and Offensive classes. In Waseem-dataset, F1-measure drops 3.7\% after reweighting highly correlated 2-grams to the Racism and Sexism classes whereas this reduction is more for Davidson-dataset. After re-weighting highly correlated 2-grams to the Hate and Offensive classes in Davidson-dataset, F1-measure drops 5.5\%. The main intuition behind this phenomenon is that both training and test sets have the same phrase distribution per class as shown in Figs~\ref{fig:bigram_waseem} and ~\ref{fig:bigram_davidson}. Due to the high correlation between specific 2-grams and a class label, reweighting the training samples results in reducing this correlation and increasing misclassification cases for the test set. Results indicate that this kind of correlation between specific words and labels in Davidson-dataset is higher than Waseem-dataset because the performance reduction is more by applying the re-weighting mechanism.

\section*{Discussion and challenges}
Although our pre-trained BERT-based model\cite{mozafari2019} has achieved promising results in terms of F1-measure on Waseem and Davidson test sets (Table \ref{bert-fine-tune}), the existing biases in data cannot be captured and measured by a test set at which there is the same biased distribution as training and validation sets. Therefore, we use a cross-domain approach to evaluate our de-biased model. Using the cross-domain approach and demonstrating the results reveals that our classifiers trained on these datasets have systematic and substantial biases where tweets written in AAE are particularly predicted as negative classes (racism, sexism, hate or offensive contents) compared with SAE (Table \ref{bias_analysis_1}). To get more insight into the differences between dialects used in tweets written in AAE and SAE, we extracted the most frequently occurred unigrams and 2-grams in both groups included in \nameref{S1_Table}. We found that there are particular words and phrases, which are more frequently used by AAE rather than SAE, and they are more related to negative classes in training datasets.

We inspected the samples in both AAE and SAE groups that are predicted as racism by applying trained classifiers with and without re-weighting mechanism. The classifier trained on Waseem-dataset without reweighting, surprisingly classifies AAE samples as racism with a higher rate than SAE (Almost 10 times). However, for both AAE-aligned and SAE-aligned groups, the number of samples assigned to racism class is very low, which can be owing to two presumptions. The first is the characteristics associated with racism samples in training data in Waseem-dataset where the majority of samples comprise religion and anti-Muslim contents, which are totally different from anti-black language used in AAE and SAE groups. The second one is mainly related to contextual knowledge derived from the pre-trained BERT model. We investigated the AAE samples assigned to racism class by trained classifier, without re-weighting mechanism, and most of them contain some racial slurs such as ``n*gga'' and ``‘b*tch'' that are contextually related to racial contents. However, after applying re-weighting mechanism these numbers of samples are reduced and result in a trade-off between AAE and SAE samples assigned to racism class and alleviating racial bias in our trained classifier with re-weighting mechanism. Although we achieve a particular reduction in racial bias included in trained classifier by applying the generalization mechanism, reweighting the training data, we believe that still some biases exist in our trained classifiers after reweighting the samples that are associated with the general knowledge of pre-trained BERT model and it should be considered as future work.

Analyzing the samples in AAE group predicted as sexism reveals that our classifier trained on training data without leveraging the re-weighting mechanism, has a high tendency to classify AAE-aligned samples containing common words in AAE language and related to feminism as sexism. However, after reducing the effect of most frequently used $n$-grams ($n=2$) in training data with applying the re-weighting mechanism, this likelihood is reduced. As Park et al.\cite{park2018} asserted the existence of gender biases in Waseem-dataset, it can be inferred that our re-weighting mechanism needs to address the gender bias in training data as long as most frequently used $n$-grams to alleviate the bias in trained model more efficiently for sexism class.

Turning to the Davidson-dataset, we observed reducing the racial bias for both Hate and Offensive classes after applying the re-weighting mechanism (Table \ref{bias_analysis_1}). Given the words associated with AAE language and highly correlated to the Hate and Offensive classes in Davidson-dataset such as ``n*gga'' and ``b*tch'' \cite{Waseem2018}, a substantially higher rate of AAE-aligned samples classified as hate and offensive than SAE-aligned can be justified; where the number of tweets containing ``n*gga'' and  ``b*tch'' in AAE-aligned samples is thirty and five times more than SAE-aligned samples. As it is noted in\cite{Waseem2018, sap2019}, these kinds of words are common in AAE dialects and used in daily conversations, therefore, it more probably will be predicted as hate or offensive when are written in SAE by associated group.

In summary, we should consider in future studies paying substantial attention to sexual and gender identities as long as dialect and social identity of the speaker in concert with highly correlated $n$-grams with the negative classes to make the bias alleviation mechanism more precise and effective. On the other hand, using pre-trained language modeling approaches such as BERT may include some general and external knowledge to the classifier, which may be a source of bias itself and it is worth further investigation.

\section*{Conclusion}
This study reveals that the benchmark datasets for hate speech and abusive language identification tasks are containing oddities that cause a high preference for classifiers to classify some samples to the specific classes. These oddities are mainly associated with a high correlation between some specific $n$-grams from a training set and a specific negative class. Employing a cross-domain evaluation approach, using the classifiers trained on these datasets, demonstrates some systematic biases in these classifiers. Therefore, we use a bias alleviation mechanism to decrease the impact of oddities in training data using a pre-trained BERT-based classifier, which is fine-tuned with a new reweighted training set. The experiments show the ability of the model in decreasing racial bias. We believe our results have made an important step towards debiasing the training classifiers for hate speech and abusive language detection tasks where the systematic bias is an intrinsic factor in hate speech detection systems. An interesting direction for future research would be to consider sexual and gender identities as long as the dialect and social identity of speakers along with $n$-grams to make the re-weighting mechanism more general and independent from training data. Furthermore, investigating the effect of samples’  weights in the compatibility function of the BERT model rather than in the classification loss function maybe improve the result. Most work has so far focused on AAE/SAE language, but it remains to be seen how our debiasing approach or any of the other prior approaches would fare in other cross-domain datasets containing different language dialects.

\section*{Supporting information}

\paragraph*{S1 Fig.}
\label{S1_Fig}
{\bf Sentence embeddings extracted from 12 layers of the pre-trained BERT model before and after fine-tuning with training and validation sets of Davidson-dataset.}
\begin{figure}[]
\centering
\begin{adjustwidth}{-1.25in}{0in}
\begin{subfigure}[b]{1.1\textwidth}
  \includegraphics[width=1\linewidth]{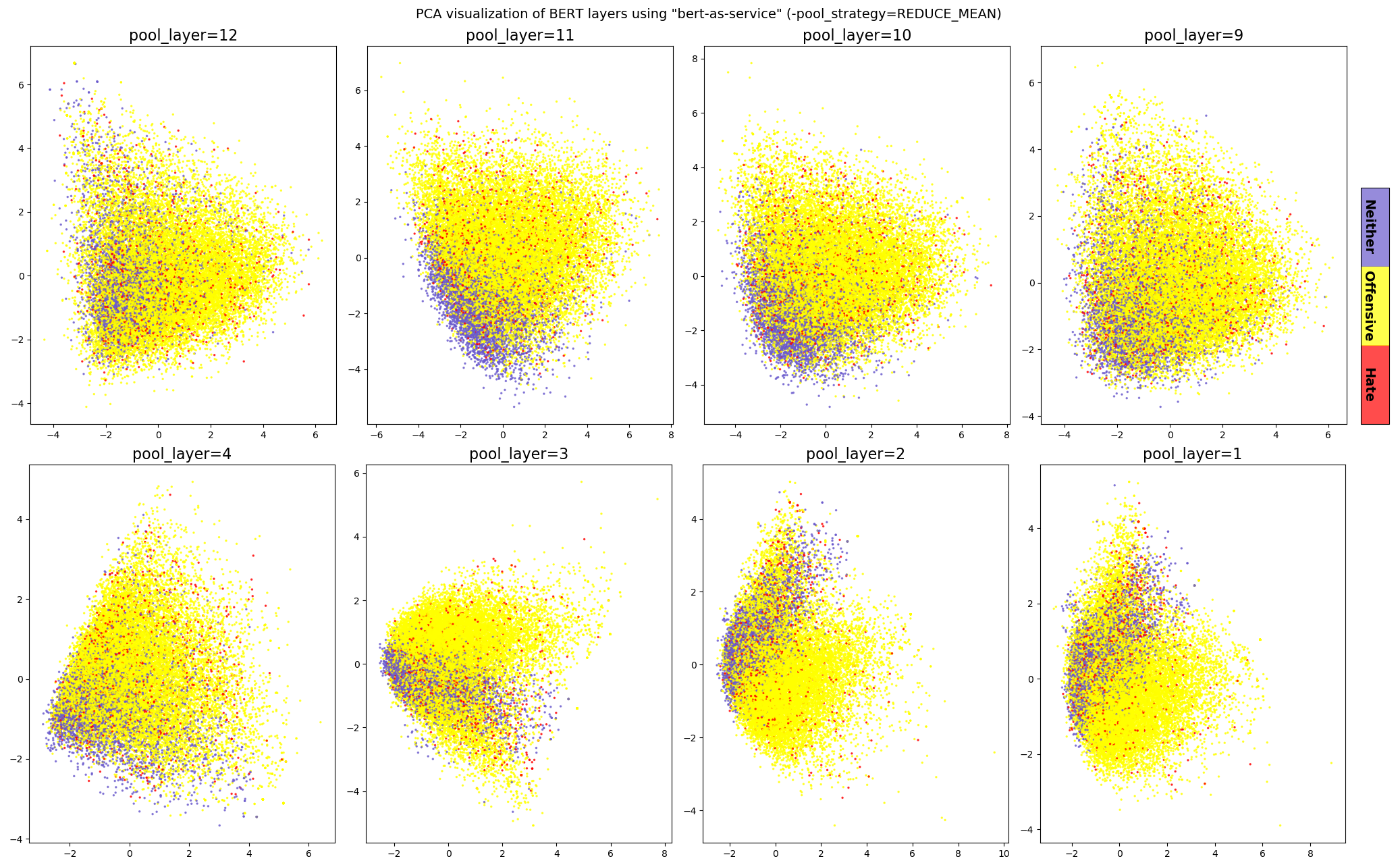}
   \caption{\centering Before fine-tuning}
   \label{fig:d_initial} 
\end{subfigure}
\begin{subfigure}[b]{1.1\textwidth}
   \includegraphics[width=1\linewidth]{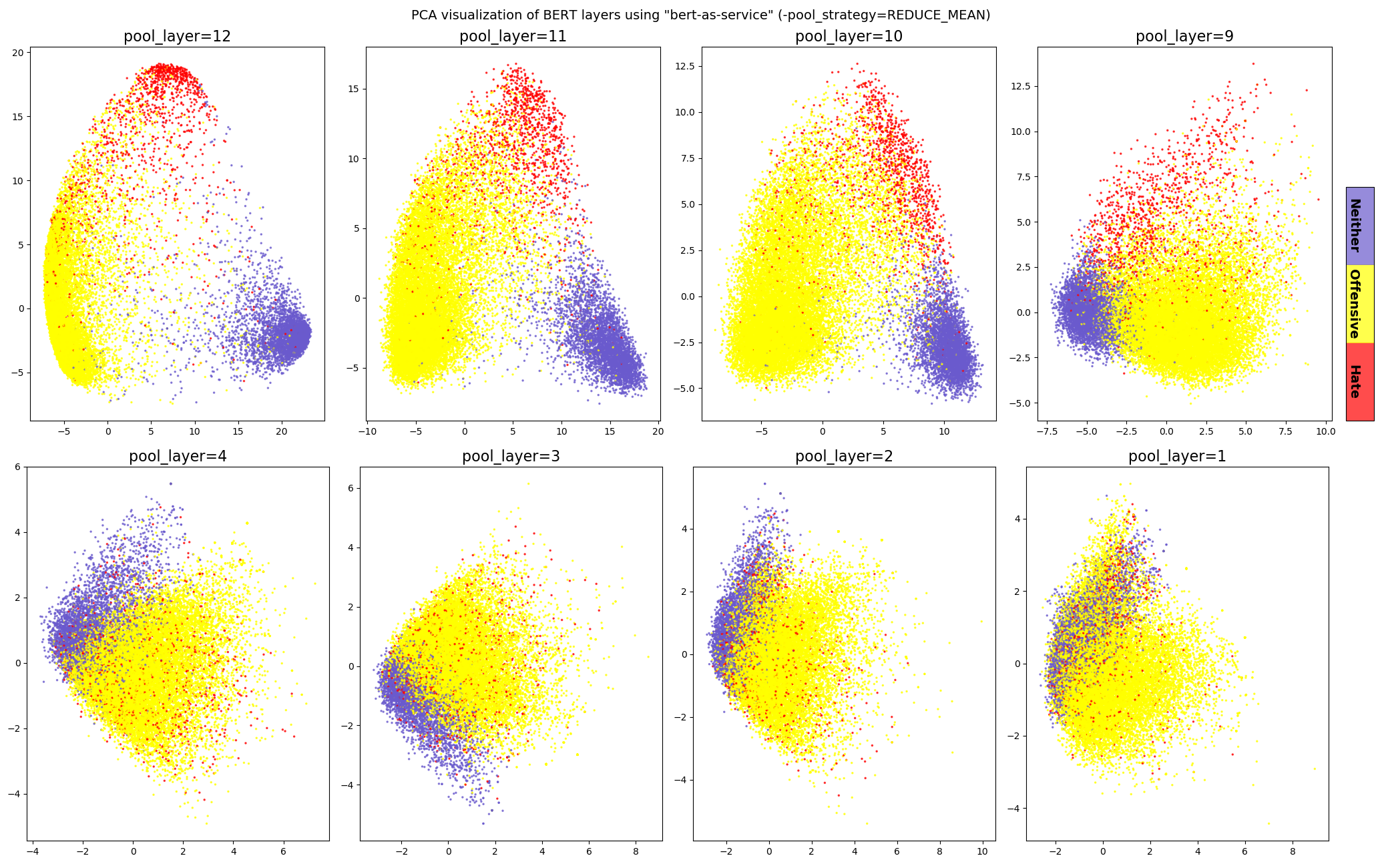}
   \caption{\centering After fine-tuning}
   \label{fig:d_enrich}
\end{subfigure}
\caption[Two numerical solutions]{\textbf{Davidson-samples' embeddings alanysis before and after fine-tuning.} To investigate the impact of information included in different layers of BERT, sentence embeddings are extracted from all the layers of the pre-trained BERT model before (a) and after (b) fine-tuning, using bert-as-service tool. Embedding vectors of size 768 are visualized to a two-dimensional visualization of the space of all Davidson-dataset samples using PCA method. For sake of clarity, we just include visualization of the first 4 layers (1-4), which are close to the training output, and the last 4 layers (9-12), which are close to the word embedding, of the pre-trained BERT model before and after fine-tuning.}
\end{adjustwidth}
\end{figure}

\paragraph*{S1 Table.}
\label{S1_Table}
{\bf Top 20 unigrams and 2-grams highly correlated with AAE and SAE languages and the number of occurrences.} Extracting unigrams and 2-grams that occur most frequently in tweets written by AAE and SAE groups, shows that some particular phrases such as ``n*gga'', ``b*tch'', ``sh*t'', ``f*ck\_w*t'', ``**s\_n*gga'', etc., are common in AAE dialects and are highly correlated with negative classes (Racism, Sexism, Hate and Offensive) in hate and offensive datasets.
\begin{table}[h]
\begin{adjustwidth}{-2.2in}{0in}
\caption{{\bf Top 20 unigrams and 2-grams highly correlated with AAE and SAE languages.} Extracting unigrams and 2-grams that occur most frequently in tweets written by AAE and SAE groups, shows that some particular phrases such as ``n*gga'', ``b*tch'', ``sh*t'', ``f*ck\_w*t'', ``**s\_n*gga'', etc., are common in AAE dialects and are highly correlated with negative classes (Racism, Sexism, Hate and Offensive) in hate and offensive datasets.}
\begin{tabular}{@{}c|l|l@{}}
\toprule
\multicolumn{1}{l|}{} & \multicolumn{1}{c|}{\textbf{unigrams}}                                                                                                                                                                                                                                                                              & \multicolumn{1}{c}{\textbf{2-grams}}                                                                                                                                                                                                                                                                                                                                                                                                                                                                                                                                                                                            \\ \midrule
\textbf{AAE-aligned}  & \begin{tabular}[c]{@{}l@{}}(lol, 726); (sh*t, 653); (u, 574); (get, 528); \\ (like, 504); (got, 483); (n*gga, 450); (**s, 428); \\ (im, 366); (f*ck, 314); (go,  312); (know, 291); \\ (b*tch, 290); (n*ggas, 285); (bout, 272); (need, 264); \\ (good, 254); (back, 232); (love, 223);(w*t, 218)\end{tabular}     & \begin{tabular}[c]{@{}l@{}}(good\_morning, 50); (feel\_like, 38); (sh*t\_sh*t, 32); \\ (go\_sleep, 31); (f*ck\_w*t, 30); (talking\_bout, 27); \\ (talkin\_bout, 26); (look\_like, 25); (wanna\_go, 23);\\ (last\_night, 22); (yo\_**s, 22); (u\_got, 21); \\ (gotta\_get, 19); (worried\_bout, 18); (go\_back, 17); \\ (**s\_n*gga, 17); (real\_n*gga, 16); (give\_f*ck, 15); \\ (lil\_n*gga, 14); (aint\_sh*t, 14); (sh*t\_like, 13)\end{tabular}                                                                                                                                                                              \\ \midrule
\textbf{SAE-aligned}  & \begin{tabular}[c]{@{}l@{}}(like, 574); (get, 475); (go, 407); (love, 372);\\ (good, 361); (one, 339); (day, 311); (time, 282);\\ (know, 271); (night, 260); (lol, 248); (today, 246); \\ (really, 236); (back, 231); (right, 231); (people, 228);\\ (see, 226); (got, 212); (life, 184); (come, 181)\end{tabular} & \begin{tabular}[c]{@{}l@{}}(last\_night, 59); (feel\_like, 55); (let\_us, 34);\\ (wish\_could, 26); (go\_home, 25); (go\_back, 24);\\ (best\_friend, 24); (wanna\_go, 24); (need\_get, 24);\\ (wait\_see, 22); (thank\_god, 20); (looks\_like, 20);\\ (good\_day, 20); (first\_time, 20); (good\_night, 19);\\ (fall\_asleep, 18); (good\_luck, 17); (come\_back, 15); \\ (great\_day, 15);(high\_school, 15);(holy\_sh*t, 13)\\
\end{tabular} \\ \bottomrule
\end{tabular}
\end{adjustwidth}
\end{table}


%
%
%
\newpage
\bibliographystyle{plos2015}

\end{document}